\documentclass[preprint]{revtex4}

\begin{document}

\title{Decaying $\Lambda$ cosmology with varying G}

\author{Saulo Carneiro}

\affiliation{Instituto de F\'{\i}sica, Universidade Federal da
Bahia, 40210-340, Salvador, BA, Brazil}

\begin{abstract}
We study a uniform and isotropic cosmology with a decaying vacuum
energy density, in the realm of a model with a time varying
gravitational ``constant". We show that, for late times, such a
cosmology is in accordance with the observed values of the
cosmological parameters. In particular, we can obtain the observed
ratio between the matter density and the total energy density,
with no necessity of any fine tuning.
\end{abstract}

\maketitle

\section{Introduction}

The advent of precise observational cosmology and the recent
discovery of a small cosmological constant $\Lambda$ have
reinforced two theoretical problems that have remained open
throughout the years \cite{GRF}. The first is the absolute value
of $\Lambda$, $122$ orders of magnitude smaller than the value
expected for the Planck time. The second is why the vacuum energy
density has, at the present time, the same order as the matter
density. This second problem is usually known as the cosmic
coincidence problem.

A possible explanation for the small value of $\Lambda$ can be
based on the idea that the vacuum energy density is not constant,
but decays as the universe expands \cite{Wu}-\cite{Narlikar}.
Nevertheless, this does not explain the cosmic coincidence, except
by means of a fine tuning of initial conditions.

In the present contribution, we will consider a FLRW, spatially
flat, decaying $\Lambda$ cosmology, in the realm of a model where
the gravitation ``constant" G varies with time at a cosmological
scale. We will show that, in the asymptotic limit of late times,
our model admits, besides the usual de Sitter solution, three
other solutions characterized by a constant ratio between the
matter density and the total energy density. Two of them have a
decelerating expansion. The third one presents a coasting
expansion, with the universe age given by $Ht=1$, and a relative
matter density given by $\rho_m/\rho=1/3$.


\section{Decaying $\Lambda$ solutions with constant $G$}

In the flat case, the Einstein equations are given by
\begin{eqnarray}
&\rho = 3 H^2,& \\ &\dot{\rho}+3H(\rho + p)=0,&
\end{eqnarray}
where $H=\dot{a}/a$ is the Hubble parameter.

We will consider a twofold energy content, formed by dust matter
with energy density $\rho_m$, and by a vacuum term with equation
of state $p_{\Lambda}=-\rho_{\Lambda}$. Then we have, for the
total energy and pressure,
\begin{eqnarray}
\rho&=&\rho_m+\rho_{\Lambda},\\ p&=&-\rho_{\Lambda}.
\end{eqnarray}

Substituting (1), (3) and (4) into (2), we obtain
\begin{equation}
2\dot{H}+\rho_m=0.
\end{equation}

If we suppose that, for late times, $H$ and $\rho_m$ falls
monotonically with the scale factor $a$, we can expand them in
negative power series of $a$. Taking the dominant terms, we then
have
\begin{eqnarray}
H&=&\beta/a^k,
\\
\rho_m&=&\gamma/a^n,
\end{eqnarray}
where $n$ and $k$ are positive integers.

Substituting into (5) leads to
\begin{equation}
\gamma a^{-n}-2k\beta^2a^{-2k}=0.
\end{equation}

For the above equation to be valid for any (large) value of $a$,
we must have
\begin{eqnarray}
n &=& 2k, \\ \gamma&=&2\beta^2k.
\end{eqnarray}

Now, it is straightforward to show that
\begin{equation}
\Omega_m \equiv \rho_m/\rho = 2k/3.
\end{equation}

Since $\Omega_m \le 1$, it follows that $k=0,1$. In the case
$k=0$, we have
\begin{eqnarray}
H &=& \beta, \\ \rho_m&=&0.
\end{eqnarray}
This solution corresponds to a de Sitter universe, with a constant
vacuum energy density.

In the case $k=1$, it follows that
\begin{eqnarray}
a&=&\beta t,
\\ H t &=& 1, \\ \rho_m/\rho &=& 2/3.
\end{eqnarray}

As discussed in \cite{Alcaniz,Alcaniz2}, among the solutions of
the form $a\propto t^n$, the best fitting of supernova and radio
sources observations is obtained in the coasting case, $a\propto
t$. On the other hand, the relation $Ht=1$ gives the best
estimation for the universe age. Nevertheless, we know that a
relative matter density equal to $2/3$ is not consistent with the
observed amount of visible and dark matter.


\section{Decaying $\Lambda$ solutions with varying $G$}

In the realm of a varying gravitational ``constant", we will still
consider the ansatz
\begin{eqnarray}
\rho = \frac{3H^2}{8\pi G}, \\ \dot{\rho}+3H(\rho + p)=0,
\end{eqnarray}
where we have re-introduced the factor $8\pi G$.

The continuity equation (18) does not depend on the varying or
constant character of $G$, being just an expression of energy
conservation. As far as relation (17) is concerned, it is valid
today, and we will assume that it is valid for any late time.

As to the variation law for $G$, let us take the
Eddington-Weinberg empirical relation \cite{GS}
\begin{equation}
G \approx \frac{H}{m^3} = \frac{H}{8\pi \lambda},
\end{equation}
where $m$ has the order of the pion mass, and the constant
$\lambda$ was introduced for convenience. Once again, this
relation is valid today, and we will assume that it is valid for
any late time.

Substituting (19) into (17), we have
\begin{equation}
\rho = 3\lambda H.
\end{equation}

For the total energy density and pressure, we will use again
\begin{eqnarray}
\rho &=& \rho_m + \rho_{\Lambda}, \\ p &=& -\rho_{\Lambda}.
\end{eqnarray}

Carrying (20-22) into the continuity equation (18), we obtain
\begin{equation}
\lambda \dot{H} + \rho_m H = 0.
\end{equation}

Let us expand again $H$ and $\rho_m$ in negative powers of $a$,
and take the leading terms
\begin{eqnarray}
H &=& \beta/a^k, \\ \rho_m &=& \gamma/a^n.
\end{eqnarray}

Substituting into (23) leads to
\begin{equation}
\gamma a^{-n} - \lambda \beta k a^{-k} = 0.
\end{equation}

Now, the possible solutions are given by
\begin{eqnarray}
k&=&n, \\ \gamma &=& \lambda \beta n.
\end{eqnarray}

It is easy to show that this leads to
\begin{equation}
\rho_m/\rho = n/3,
\end{equation}
which means that, now, $n=0,1,2,3$.

For $n=0$, we have
\begin{eqnarray}
H&=&\beta, \\ \rho_m&=&0, \\ G&=&\frac{\beta}{8\pi \lambda}.
\end{eqnarray}
That is, the usual de Sitter solution, with $\Lambda$ and $G$
constant.

For the other values of $n$, we obtain
\begin{eqnarray}
a &=& (n\beta t)^{1/n}, \\ H t&=&1/n, \\ \rho_m/\rho &=& n/3.
\end{eqnarray}

For $n=2,3$, we have decelerating solutions, leading to a too
young universe and a too high matter density. For $n=1$, it
follows that
\begin{eqnarray}
a&=&\beta t, \\ H t &=& 1, \\ \rho_m/\rho &=& 1/3.
\end{eqnarray}

Therefore, we re-obtain a coasting expansion, with an acceptable
age for the universe. Moreover, this time the corresponding
relative matter density matches surprisingly well the observed
value.


\section{Concluding remarks}

To conclude, let us obtain the variation rate of $G$, and the rate
of matter production in this model.

For $G$ we have used the evolution law
\begin{equation}
G = \frac{H}{8\pi \lambda},
\end{equation}
which leads to the relative variation rate
\begin{equation}
\dot{G}/G = -(1+q)H,
\end{equation}
where $q=-a\ddot{a}/\dot{a}^2$ is the deceleration parameter.

For $n=1$, we have $q=0$, and so
\begin{equation}
\dot{G}/G = -H.
\end{equation}

For the rate of matter production (coming from the decaying vacuum
energy), we have (for $n=1$)
\begin{equation}
\frac{1}{a^3}\frac{d}{dt}(\rho_m a^3)= 2\rho_m H.
\end{equation}

Finally, a note on the variation of the vacuum energy in the
present context. It is possible to show that the vacuum energy
density and the corresponding cosmological ``constant" are given
by
\begin{eqnarray}
\rho_{\Lambda} &=& (3-n)\lambda H = \frac{(3-n)}{8\pi}\;m^3H, \\
\Lambda &=& 8\pi G \rho_{\Lambda} = (3-n) H^2,
\end{eqnarray}
leading to present values in agreement with observation.

The first equation is in accordance with a recent derivation by
Sch\"utzhold \cite{Ralf}, based on quantum field calculations in
an expanding background. The second equation agrees with an ansatz
originally proposed by Chen and Wu \cite{Wu}, and generalized
later by Carvalho, Lima and Waga \cite{Waga}, and by Alcaniz and
Maia \cite{Maia}. Note that, in the realm of a constant $G$, those
two equations would be incompatible to each other.

${}$

I would like to thank Ralf Sch\"utzhold for useful discussions.



\begin{thebibliography}{}

\bibitem{GRF} S. Carneiro, gr-qc/0305081, to appear in the GRF2003 Special Issue of
IJMPD; gr-qc/0206064.

\bibitem{Wu} W. Chen and Y.S. Wu, Phys. Rev. D {\bf 41}, 695 (1990).

\bibitem{Waga} J.C. Carvalho, J.A.S. Lima and I. Waga, Phys. Rev.
D {\bf 46}, 2404 (1992).

\bibitem{Maia} J.S. Alcaniz and J.M.F. Maia, Phys. Rev. D {\bf
67}, 043502, 2003.

\bibitem{Novello} M. Novello, J. Barcelos-Neto and J.M. Salim, Class. Quant.
Grav. {\bf 18}, 1261 (2001); {\bf 19}, 3107 (2002).

\bibitem{JJ} M.V. John and K.B. Joseph, Phys. Rev. D {\bf 61},
087304 (2000);

\bibitem{Narlikar} M.V. John and J.V. Narlikar, Phys. Rev. D {\bf 65}, 043506
(2002).

\bibitem{Alcaniz} D. Jair, A. Dev, and J.S. Alcaniz, astro-ph/0302025;

\bibitem{Alcaniz2} A. Dev, M. Sethi, and D. Lohiya, Phys. Lett. B {\bf 504}, 207
(2001).

\bibitem{GS} G.A. Mena Marug\'{a}n and S. Carneiro, Phys. Rev. D {\bf 65},
087303 (2002);

\bibitem{Ralf} R. Sch\"utzhold, Phys. Rev. Lett. {\bf 89}, 081302
(2002).

\end{thebibliography}
\end{document}